\documentclass[twocolumn,pra, aps,showpacs]{revtex4-2}
\usepackage{subfigure}
\usepackage{psfrag}
\usepackage{dcolumn}
\usepackage{bm}
\usepackage{color}
\usepackage{latexsym}
\usepackage{epstopdf}
\usepackage[english]{babel}
\usepackage{latexsym}
\usepackage{natbib}
\usepackage{graphicx}
\usepackage{epsf}
\usepackage{amsmath}
\usepackage{amssymb}
\usepackage{amsfonts}
\usepackage{appendix}
\usepackage{lipsum, babel}
\usepackage{soul}
\usepackage{hyperref}
\hypersetup{colorlinks=true,citecolor=myblue,linkcolor=blue,urlcolor=myblue}

\usepackage{siunitx}
\usepackage{xcolor}
\usepackage[draft,inline,nomargin,index]{fixme}
\usepackage[normalem]{ulem}

\fxsetup{theme=color,mode=multiuser}

\FXRegisterAuthor{ofer}{aOfer}{\color{brown} Ofer} 
\FXRegisterAuthor{bankim}{abankim}{\color{red}  Bankim}
\definecolor{mygrey}{gray}{0.35}
\definecolor{myblue}{rgb}{0.2,0.2,0.8}
\definecolor{myzard}{cmyk}{0,0,0.05,0}
\definecolor{mywhite}{rgb}{1,1,1}
\definecolor{myred}{rgb}{1,0.,0.3}
\definecolor{myblack}{rgb}{0,0,0}

\def\min{\mathord{\rm min}}

\newcommand{\blue}{\color{blue}}

 \newcommand{\ket}[1]{|#1\rangle}
 \newcommand{\bra}[1]{\langle #1|}

\usepackage{dsfont}

\makeatletter

\renewenvironment{widetext@grid}{%
  \par\ignorespaces
  \setbox\widetext@top\vbox{%
   \vskip15\p@
   \hb@xt@\hsize{%
    \leaders\hrule\hfil
    \vrule\@height6\p@
   }%
   \vskip6\p@
  }%
  \setbox\widetext@bot\hb@xt@\hsize{%
    \vrule\@depth6\p@
    \leaders\hrule\hfil
  }%
  \onecolumngrid
  \let\set@footnotewidth\set@footnotewidth@ii
}{%
  \par
  \twocolumngrid\global\@ignoretrue
  \@endpetrue
}%

\makeatother
\begin{document}
\raggedbottom
%\maketitle

\title{Quantum CNOT Gate with Actively Synchronized Photon Pairs}

\author{Haim Nakav}
\thanks{These authors contributed equally to this work}
\author{Tanim Firdoshi}
\thanks{These authors contributed equally to this work}
\author{Omri Davidson}
\author{Bankim Chandra Das}
\author{Ofer Firstenberg}
\affiliation{Physics of Complex Systems, Weizmann Institute of Science and AMOS, Rehovot 7610001, Israel}

\begin{abstract}

Controlling the synchronization of photons from probabilistic quantum sources plays a pivotal role in advancing efficient quantum information processing. We report the realization of a probabilistic entangling gate operating on actively synchronized photon pairs, using a quantum memory based on warm atomic vapor. We achieve a truth-table fidelity exceeding 85\% and demonstrate Bell-inequality violation for all four Bell states. The reduction in visibility of the Hong-Ou-Mandel (HOM) interference introduced by the storage process is identified as a primary factor limiting the gate fidelity. We derive an exact quantitative relation between HOM visibility and gate fidelity for pure single photons, which applies to all photonic gates relying on interference, regardless of the physical platform. 

\end{abstract}
%\pacs{}
\maketitle

Single-photon sources based on atomic vapor are prominent platforms for photonic quantum computing and communication, due to their simplicity and scalability \cite{Glorieux_2023}. These sources often rely on spontaneous multi-photon processes, generating high-fidelity single photons in well-defined modes \cite{MoonOE2016,Davidson_2021}. Nevertheless, the stochastic nature of these sources inherently limits their ability to generate large photonic states by multiplexing, as the success probability decreases exponentially with the photon number. To this end, integrating quantum memories with photon sources is imperative \cite{nunn2013}. In particular, atomic-vapor memories can store single photons from spontaneous parametric down-conversion (SPDC) or four-wave mixing (FWM) sources \cite{MichelbergerNunnNJP2015,Walmsley_2018,Treutlein_2022,Davidson_2023_prl}.  
These memories thus enhance the generation of photon pairs,
facilitating the efficient operation of two-photon gates.

Photonic entangling gates are most commonly implemented using linear optical components within the Knill-Laflamme-Milburn (KLM) framework \cite{KLM2001}. Being probabilistic, these gates rely on post-selection but nevertheless are proven to be sufficient for scalable quantum computation \cite{kiesel2005linear,Okamoto2005,kok2007}.  Linear-optics gates are based on two-photon interference -- namely, the Hong-Ou-Mandel (HOM) interference effect. Their fidelity is governed by the HOM interference visibility. For a pair of pure single photons in identical spatial modes, the HOM visibility reflects their temporal indistinguishability
\cite{Kambs_2018,legero2003time}. Maintaining indistinguishability becomes particularly challenging when interfacing photons from heterogeneous modules, such as a source and a memory.

\begin{figure*}[!t] %!ht % "t" means the top of the page; you can change it to "b" for bottom or "h" for here
  \centering
  \includegraphics[width=0.9\textwidth]{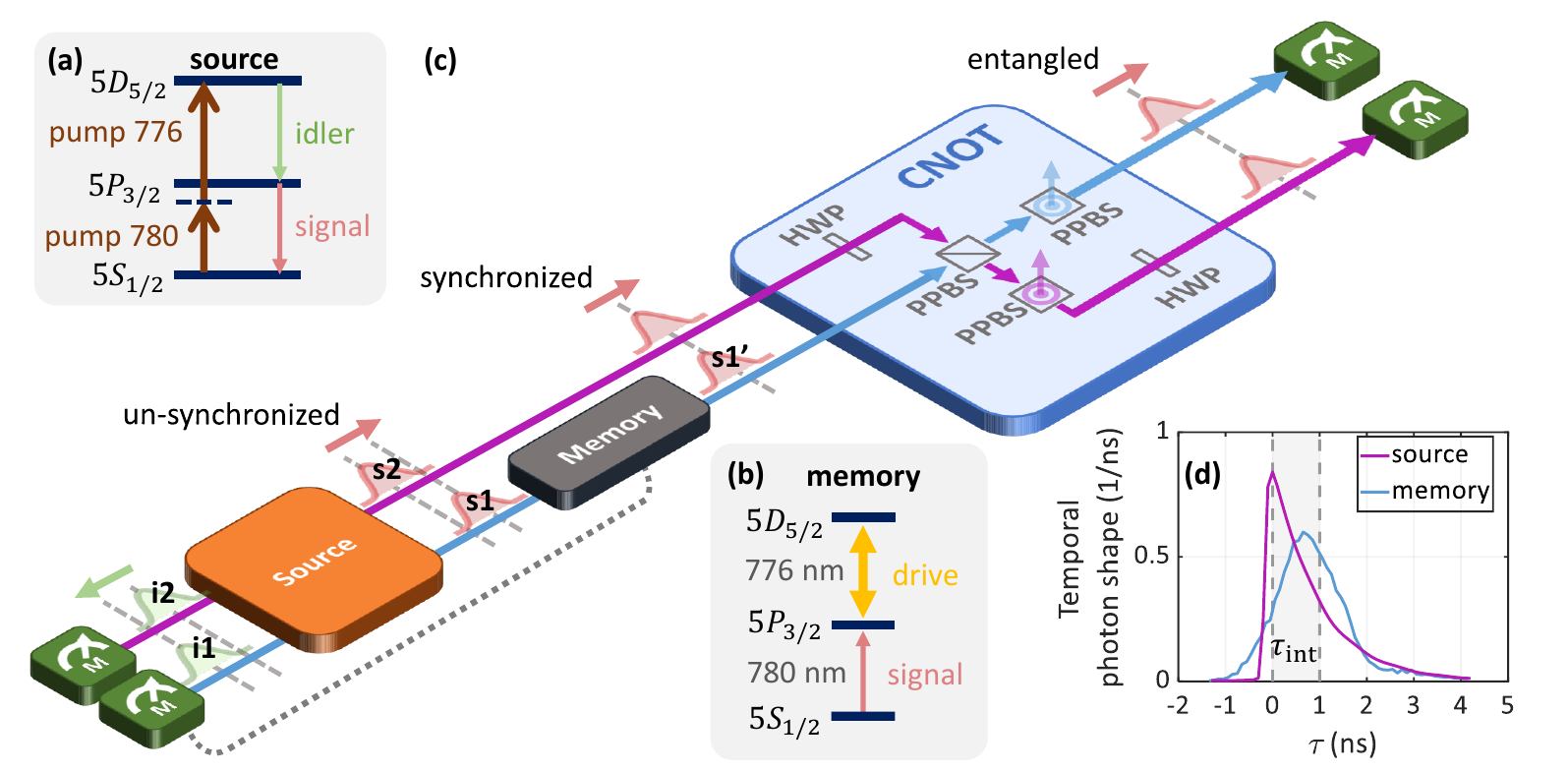}
  \caption{\textbf{Experimental scheme for a quantum CNOT gate with actively synchronized photons.} 
  \textbf{(a)} Atomic level scheme of the photon source, stochastically generating pairs of signal and idler photons in two different channels through spontaneous four-wave mixing. 
   \textbf{(b)} Atomic level scheme of the quantum memory. 
   \textbf{(c)} A signal photon (s1) is stored when detecting the idler photon (i1) and retrieved (s1$'$) when detecting the second idler photon (i2). This synchronizes s1$'$ with the second signal photon (s2), which are sent to the CNOT setup. The input qubit states are encoded using polarization wave-plates (not shown). The gate operation relies on two-photon interference on partially-polarizing beam splitters (PPBS). The outgoing qubit states are measured via polarization tomography using single-photon detectors. 
   \textbf{(d)} The temporal shapes of the synchronized photons show that the photons retrieved from the memory (blue) are temporally broader (full-width at half maximum, FWHM$=1.5~$ns) and more symmetric compared to the photons produced by the source (purple, FWHM$=0.9~$ns).
   }
  \label{fig:Experimental setup}
  
\end{figure*}

In this work, we realize a KLM-based two-photon quantum gate executing a controlled-not (CNOT) operation in the polarization basis of actively synchronized photon pairs.  
For the gate input, we synchronize two single photons by combining a stochastic single-photon source \cite{davidson2023newsource} with a quantum memory \cite{Davidson2023, Davidson_2023_prl}. The source and memory use an identical atomic level scheme in warm vapor, ensuring mutual compatibility, maximizing memory performance, and preserving high photon indistinguishability. Our fast ladder memory (FLAME) \cite{Finkelstein2018} can store nanosecond-long single photons and has negligible noise (15$\times 10^{-5}$ noise photons per retrieved photon). 
The integration of FLAME with the photon source dramatically enhances the rate of synchronized photon pairs arriving at the quantum CNOT gate, which in turn leverages their high indistinguishability for producing high-fidelity Bell states. Our experiment realizes a key step toward scalable photonic quantum computing.

\vspace{0.2\baselineskip}
\textit{Experimental setup.---} The first part of the setup is responsible for generating synchronized photon pairs. 
It includes a dual-channel single-photon source and the FLAME memory, as described in Ref.~\cite{Davidson_2023_prl}, both utilizing the ladder-type orbital-level scheme $5S_{1/2} \rightarrow 5P_{3/2} \rightarrow 5D_{5/2}$ of $^{87}$Rb, shown in Fig.~\ref{fig:Experimental setup}(a,b). The source is driven by two pump lasers and stochastically generates signal and idler photons (at 780 nm and 776 nm, respectively) along phase-matched directions in two different channels. In both channels, the detection of an idler photon heralds the generation of a signal photon. 

The present source generates signal photons at an average rate of 40,000 counts/sec per channel, with heralding efficiency of 18\% and 22.7\% for channels 1 and 2, respectively, a multi-photon component of $g^{(2)}_\mathrm{h}(0)=0.0129(1)$,
and HOM visibility of $\mathcal{V}_\mathrm{source}=0.89(2)$ [\citealp{SM}, Sec.~S1].
 A signal photon from one channel is sent to FLAME, where it is stored by overlapping it with a drive-field pulse. The present FLAME has an end-to-end efficiency of 25$\%$, a $1/e$ lifetime of nearly 100 ns, and the retrieved photons exhibit $g^{(2)}_\mathrm{h}(0)=0.028(5)$ 
 [\citealp{SM}, Sec.~S2]. The retrieval of the stored photon using another drive pulse is synchronized with the generation of a signal photon in the second source channel. The two corresponding idler photons are used for triggering, as illustrated in Fig.~\ref{fig:Experimental setup}c.  
 
Figure \ref{fig:Experimental setup}d presents the temporal profiles of the synchronized photons. Due to the finite bandwidth of FLAME, acting as a spectral filter, the retrieved photons are 65\% longer and temporally more symmetric than those emitted from the source. 
The relative timing between the two photons, indicated in Fig.~\ref{fig:Experimental setup}d, is tuned to maximize their HOM visibility  $\mathcal{V}$, reaching
$\mathcal{V}=0.79(2)$ [\citealp{SM}, Sec.~S3].

The synchronized photons are directed via fibers to the second part of the setup, comprising qubit state preparation, the CNOT gate, and measurement. 
We use the linear polarization states $\ket{\mathrm{H}} \equiv \ket{0}$ and $\ket{\mathrm{V}} \equiv \ket{1}$ as the CNOT's basis, that is, the \textit{target} qubit flips between $\ket{\mathrm{V}}\leftrightarrow\ket{\mathrm{H}}$ when the \textit{control} qubit is in $\ket{\mathrm{V}}$. The photon arriving from the memory acts as the control, while that arriving directly from the source serves as the target. We prepare the initial qubit states using waveplates. 

Our CNOT implementation, based on Ref.~\cite{He2013,zhai2022quantum}, is shown in Fig.~\ref{fig:Experimental setup}c. The control and target photons impinge on a partially polarizing beam splitter (PPBS) having 1 and 1/3 transmission for the horizontal and vertical polarizations, respectively. The HOM interference on this PPBS generates non-classical correlations between the two qubits. 
Additional PPBSs along each path, oriented perpendicular to the first PPBS, balance the output amplitudes for all polarization configurations.
Two half-wave plates flanking the PPBSs in the target qubit path are oriented at $22.5^\circ$ relative to the horizontal axis and act as Hadamard gates, transferring the conditional phase shift to a full CNOT operation. Two of the four output ports are connected to detection modules, enabling full-state analysis using waveplates, polarizing beam splitters, and four single-photon detectors.

 \begin{figure*}[t]
    \centering
    \includegraphics[width=0.95\textwidth]{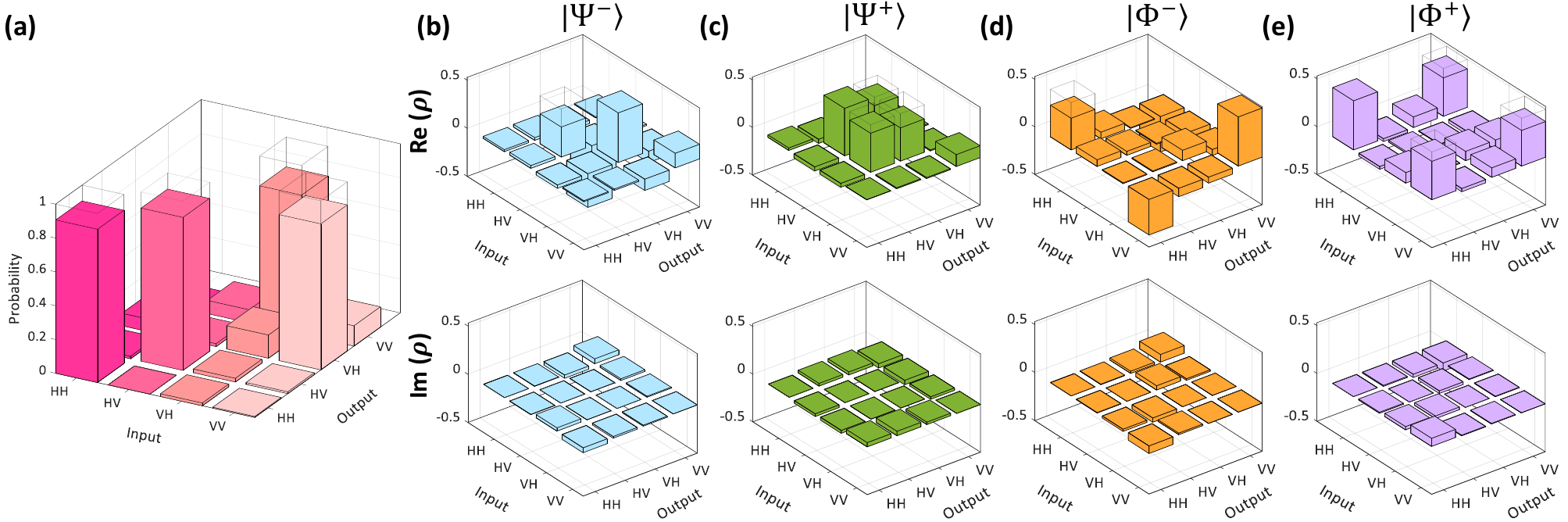}
    \caption{\textbf{CNOT gate performance. (a)} Measured truth table for the linear H/V polarization basis. \textbf{(b-e)} Reconsructred density matrices for the four Bell states.  Solid bars are experimental data, transparent region represents the ideal state.}
    \label{fig:fig2}
\end{figure*}

In an ideal system, two photons are detected in the monitored output ports in 1 out of every 9 attempts, signaling a successful gate operation with unit fidelity. In our system, imperfections in the PPBSs reduce the success probability to 1 in 9.7, with a negligible effect ($\sim 0.8\%$) on the gate fidelity.
See [\citealp{SM}, Sec.~S4] for further details on the gate setup and operation.

\vspace{0.2\baselineskip}
\textit{Results.---} The synchronization procedure increases the heralded photon-pair coincidence rate by a factor of 19 compared to the source's accidental pair rate $R_\textrm{unsync}=0.77$/sec. The resulting rate of synchronized, heralded photon pairs arriving at the gate is $R_\textrm{sync}=14.7$/sec. While the relative timing between the control and target photons is fixed, we set by post-selection the duration of the integration window $\tau_\textrm{int}$ included in the analysis, as illustrated in  Fig.~\ref{fig:Experimental setup}d. Reducing $\tau_\textrm{int}$ increases the HOM visibility $\mathcal{V}$, enabling a tunable tradeoff between gate efficiency and fidelity \cite{wang2025}. 

We begin by characterizing the gate performance for $\tau_\textrm{int}=1$ ns, which accepts 56\% of the photons and yields  $\mathcal{V}=0.90(2)$.
Figure \ref{fig:fig2}a shows the ZZ truth table, obtained by preparing the incoming qubits in the four basis states of the linear polarization basis H/V and measuring the coincidence counts between different detector pairs in the same basis (H/V).  The fidelity of this truth table is $F_\textrm{ZZ}=0.86(1)$. 
The two other, complementary truth tables are XX and YY, obtained by preparing the input qubits in the D/A basis $\ket{\mathrm{D/A}}=\frac{1}{\sqrt{2}}(\ket{\mathrm{H}} \pm \ket{\mathrm{V}})$ or in a combination of D/A and H/V, and measuring the output in either the D/A or R/L basis $\ket{\mathrm{R/L}}=\frac{1}{\sqrt{2}}(\ket{\mathrm{H}} \pm i \ket{\mathrm{V}})$  {\blue{\cite{altepeter2005photonic}}}. The measured truth-table fidelities are 
$F_\textrm{XX}=0.898(4)$ and $F_\textrm{YY}=0.896(5)$. These fidelities set upper and lower bounds on the gate's process fidelity $F_\textrm{ZZ}+F_\textrm{XX}-1\leq F_\textrm{process} \leq \min[F_\textrm{ZZ},F_\textrm{XX}]$, where $F_\textrm{process}>0.5$ guarantees generation of entanglement 
\cite{Okamoto2005}. We obtain $0.76\leq F_\textrm{process} \leq 0.86$, crossing that threshold. 

\begin{figure}[!t]
    \centering
    \includegraphics[width=0.94\columnwidth]{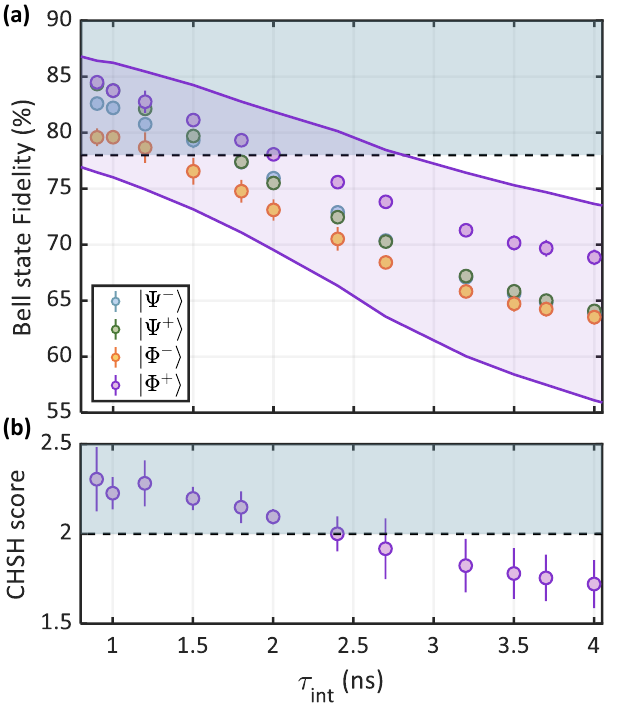}
    \caption{\textbf{Dependence of fidelity on the duration of the integration window $\tau_\textrm{int}$. (a)} Fidelity of all four Bell states.  The purple lines mark the upper and lower limits on the process fidelity, obtained from the truth-table fidelities. 
    \textbf{(b)} CHSH score of the Bell state $\ket{\Phi^{+}}$. The dashed line in both panels draws the threshold above which (a) Bell's inequality breaks, and (b) CHSH inequality breaks.
    }
    \label{fig:fig3}
\end{figure}

We use the gate to generate the four Bell states by preparing the corresponding input states, and we characterize the output using quantum state tomography. The density matrix of each output state is reconstructed from 144 coincidence count measurements (36 measurement bases, four detectors) using a maximum likelihood algorithm \cite{altepeter2005photonic,james2001measurement}. The reconstructed density matrices of $\ket{\Psi^{\mp}}=(\ket{\mathrm{HV}}\mp\ket{\mathrm{VH}})/\sqrt{2}$ and $\ket{\Phi^{\mp}}=(\ket{\mathrm{HH}}\mp\ket{\mathrm{VV}})/\sqrt{2}$ are shown in Figs.~\ref{fig:fig2}b-e, and the corresponding state fidelities are
0.822(5), 0.838(3), 0.796(2), and 0.838(7),
respectively. All fidelities are above the threshold of $(2+3\sqrt{2})/8\approx0.78$ for violating Bell's inequality \cite{kiesel2005linear}. As a complementary check, we perform a Clauser-Horne-Shimony-Holt (CHSH) measurement on the $\ket{\Phi^{+}}$ state and obtain the score $S=2.23(9)$, violating the CHSH inequality $|S|\le2$ by 2.5 standard deviations.

We now examine the dependence of the gate performance on the integration window duration $\tau_\textrm{int}$. Figure \ref{fig:fig3}a presents the measured fidelities of the four Bell states, along with the upper and lower bounds of $F_\textrm{process}$. The corresponding CHSH score for $\ket{\Phi^{+}}$ state is plotted in Fig.~\ref{fig:fig3}b. We find that the gate can generate entanglement ($F_\textrm{process}>0.5$) up to $\tau_\textrm{int}\approx4$ ns, which accepts 88$\%$ of the photons. However, as $\tau_\textrm{int}$ increases, all figures of merit drop below the thresholds for violating the Bell or CHSH inequalities. The degradation is primarily due to increasing temporal mismatch between the input photons, along with elevated multiphoton and background noise contributions at longer $\tau_\textrm{int}$. 

To isolate the effect of the memory---which alters the photon pulse shape relative to that of the source---we operated the gate without synchronization, \textit{i.e.}, using heralded, accidental photon pairs arriving directly from the source. 
Because accidental photon pairs occur at a lower rate, achieving comparable statistics requires approximately 19 times longer acquisition time. We therefore limited our measurement to the ZZ and XX truth tables, using only 10 times longer acquisition time. For an integration window of $\tau_\textrm{int}$ = 1 ns, we obtained 
$F^\textrm{unsync}_\textrm{ZZ}=F^\textrm{unsync}_\textrm{XX}=0.90(1)$, setting the bounds for the process fidelity to $0.81\leq F^\textrm{unsync}_\textrm{process} \leq 0.90$ [\citealp{SM}, Fig.~S8]. The storage and retrieval process therefore accounts for approximately a 5\% reduction in fidelity.

\begin{figure}[!t]
    \centering
    \includegraphics[width=\columnwidth]{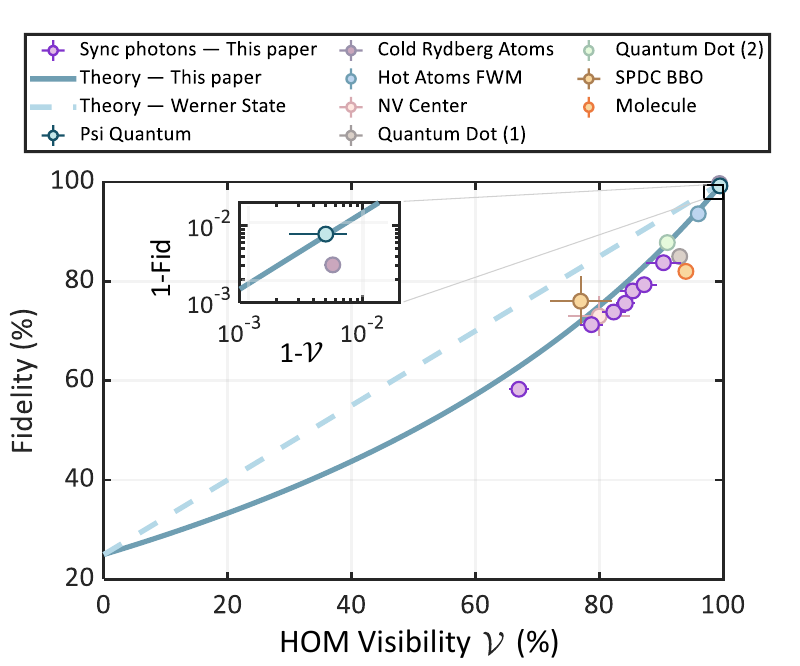}
    \caption{\textbf{Relation between state fidelity and HOM visibility} in reported implementations of HOM-based entangling gates (circles) compared to our model prediction Eq.~(\ref{eq:Fid}) (solid line, with $\eta=\mathcal{V}$). Our data points (purple) correspond to different integration-window durations. Each remaining data point represents a distinct experimental platform: cold Rydberg atoms \cite{shi2022high}, Psi Quantum (SPDC) \cite{psiquantum2025manufacturable}, Quantum dot (1) \cite{zhai2022quantum}, Quantum dot (2) \cite{li2021heralded}, NV Center \cite{bernien2013heralded}, SPDC \cite{zeuner2018integrated}, hot-atom FWM \cite{park2019polarization}, and a single molecule ($\mathcal{V}$ measured, fidelity inferred from CHSH score) \cite{rezai2019polarization}. For comparison, the fidelity predicted by the commonly assumed Werner-state model \cite{mivcuda2014process,chen_wernerstate_2024},  $F_\mathrm{Werner}(\mathcal{V})$, is shown as a dashed line. The inset zooms in on the high-visibility, high-fidelity regime.
    }
    \label{fig:fig4}
\end{figure}

\textit{Dependence of fidelity on HOM visiblity.---} 
We derive below an analytic expression relating the gate fidelity 
to the HOM interference visibility $\mathcal{V}$. We limit our model to single pure input photons, \textit{i.e.}, assuming no noise or multi-photon components at the input. Under this assumption, the HOM visibility $\mathcal{V}$ is equal to the indistinguishability $\eta$ of the input photons \cite{mosley2008heralded}.
We briefly outline the derivation here; see [\citealp{SM}] (Sec.~S5) for details. 

We denote the gate's input ports, connected to the source and memory, as `s' and `m', and the output ports as `p' and `q'. 
To simplify the derivation, we explicitly model only the temporal contribution to the indistinguishability,
\begin{equation}
\eta=\left|\int dt u^*_\mathrm{m}(t) u_\mathrm{s}(t) \right|^2\le 1~, \label{eq:eta}
\end{equation}
where $u_x(t)$ denotes the temporal profile of the photon at port $x\in\{\mathrm{m,s,p,q}\}$. Throughout, all time integrals $\int dt$ are taken over the chosen integration window, and the profiles $u_x(t)$ are normalized such that $\int dt |u_x(t)|^2=1$. Our results are nevertheless applicable to all forms of indistinguishability, if they are incorporated into $\eta$.

As a representative example, we consider the generation of the Bell state $\ket{\Phi^{+}}$ from the input state 
    $\ket{\psi_\mathrm{in}}= u_\mathrm{m}(t_\mathrm{m}) \ket{\mathrm{D}}_\mathrm{m}  u_\mathrm{s}(t_\mathrm{s})\ket{\mathrm{H}}_\mathrm{s}.$
Here, $u_x(t_x) \ket{P}_x$ is a shorthand notation describing a single-photon pulse with temporal profile $u_x(t)$ at port $x\in\{\mathrm{m,s,p,q}\}$ and polarization $P\in\{\mathrm{H,V,D,A}\}$. It is formally given by 
$u_x(t_x) \ket{P}_x \equiv \int dt~ u_x(t) a^\dagger_{x,P}(t)\ket{0}$,
where $a^\dagger_{x,P}(t)$ is the photon creation operator at time $t$. 

Since the gate relies on interference, we define $f(t,t')=u_\mathrm{m}(t) u_\mathrm{s}(t')$ and the (anti)symmetrized components $f_\pm(t,t')=[f(t,t')\pm f(t',t)]/2$. The input state can then be rewritten as
\begin{equation}
    \ket{\psi_\mathrm{in}}= f(t_\mathrm{m},t_\mathrm{s})\ket{\mathrm{DH}}_\mathrm{m,s}.
    \label{psi_in}
\end{equation}
The gate operation consists of a Hadamard transformation on port `s', followed by the three PPBS elements, and then a Hadamard on `q'. Post-selecting on events in which both photons exit through the monitored output ports, and assuming no other losses or imperfections, the output state becomes  $\ket{\psi_\mathrm{out}}=$
\begin{equation}
     \frac{f(t_\mathrm{p},t_\mathrm{q})\ket{\mathrm{HH}}_\mathrm{p,q}+2f_-(t_\mathrm{p},t_\mathrm{q})\ket{\mathrm{VH}}_\mathrm{p,q}+f(t_\mathrm{q},t_\mathrm{p})\ket{\mathrm{VV}}_\mathrm{p,q}}{\sqrt{4-2\eta}} . 
     \label{psi_out}
\end{equation}
Introducing four orthonormal Bell-like states $\ket{\tilde\Phi^\pm}=f_\pm(t_\mathrm{p},t_\mathrm{q})(\ket{\mathrm{HH}}_\mathrm{p,q}\pm\ket{\mathrm{VV}}_\mathrm{p,q})/\sqrt{1\pm\eta}$ and $\ket{\tilde\Psi^\pm}=f_-(t_\mathrm{p},t_\mathrm{q})(\ket{\mathrm{HV}}_\mathrm{p,q}\pm\ket{\mathrm{VH}}_\mathrm{p,q})/\sqrt{1-\eta}$, we can express the output as
\begin{equation}
\ket{\psi_\mathrm{out}}=\sqrt{F(\eta)}\ket{\tilde\Phi^+}+\sqrt{\frac{1-\eta}{4-2\eta}}(\ket{\tilde\Phi^-}+\ket{{\tilde{\Psi}}^+}-\ket{\tilde{\Psi}^-}) ,
     \label{psi_out2}
\end{equation}
where we identify
\begin{equation}
    F(\eta) =  \left| \langle \tilde\Phi^{+} | \psi_\mathrm{out} \rangle \right|^2 =  \frac{1}{2}\cdot\frac{1+ \eta}{2-\eta}
    \label{eq:Fid}
\end{equation}
as the fidelity for generating $\ket{\Phi^{+}}$. Repeating the analysis for the other Bell states results in the same expression Eq.~(\ref{eq:Fid}) for the generation fidelity. Given that $\mathcal{V}=\eta$ for pure single photons, we find $F(\mathcal{V})=(1+\mathcal{V})/(4-2\mathcal{V})$.

Alternatively, the same result can be obtained by constructing the reduced density matrix $\rho$ of the output state in Eq.~(\ref{psi_out}), by tracing over the temporal degrees of freedom,  $|\psi_\mathrm{out}\rangle\langle\psi_\mathrm{out}|$.
The fidelity is then given by $F(\eta)=\mathrm{Tr}(\rho\ket{\Phi^+}\bra{\Phi^+})$, yielding the same expression as in Eq.~(\ref{eq:Fid}). The reduced density matrices for all four Bell states, as a function of $\eta$, are provided in [\citealp{SM}] (Sec.~S5). 

Notably, these density matrices do not represent a Werner state \cite{RiedelGarding2021Jun}, as is often assumed to model imperfections. Consequently, our result differs from the Werner-state prediction $F_\mathrm{Werner}(\mathcal{V})=(1+3\mathcal{V})/4$ \cite{mivcuda2014process,chen_wernerstate_2024}.

Figure~\ref{fig:fig4} presents a direct comparison between the theoretical fidelity $F(\mathcal{V})$  and our experimental data, based on the measured gate fidelity and HOM visibility for a range of integration-window durations. The agreement is strong, with the measured fidelity falling short by only a few percent, 
partially due to imperfections in the PPBS elements.
The figure also includes experimental data from other studies \cite{shi2022high,zhai2022quantum, psiquantum2025manufacturable,li2021heralded, bernien2013heralded, hensen2015loophole, Kambs_2018, zeuner2018integrated, park2019polarization, rezai2019polarization}, further validating the generality of our model. The excellent agreement across platforms suggests that our model accurately captures the dominant error mechanism in interference-based photonic gates.

\vspace{0.5\baselineskip}
\textit{Discussion.---} 
We have demonstrated a high-fidelity linear-optical CNOT gate using actively synchronized photon pairs from a quantum memory. The gate achieves Bell-state fidelities up to 84\%, with process fidelities exceeding the entanglement threshold even at high repetition rates. We derive and experimentally validate an analytic relation between gate fidelity and HOM visibility. While the model is derived under the assumption of pure single photons---where photon indistinguishability fully determines the visibility---we find that it accurately describes experimental results across a broad range of physical implementations. In our system, the dominant sources of infidelity are temporal distinguishability and multi-photon events, with the memory accounting for roughly half of the total. Smaller contributions (on the order of 1\%) arise from imperfections in the gate optics, limited photon purity, and detector dark counts.

The demonstrated active synchronization enhances the usable photon-pair rate by a factor of 19, enabling full quantum state tomography and precision gate characterization with modest cost in fidelity. Looking forward, tailoring the retrieval temporal dynamics of the memory \cite{novikova2008optimal} offers a pathway to further optimize indistinguishability without sacrificing rate. Our findings establish both the practical value of memory-based synchronization and the universal predictive power of Eq.~(\ref{eq:Fid}) for high-fidelity photonic quantum gates across diverse platforms.

\textit{Acknowledgements.---}
We acknowledge financial support from the Israel Science Foundation (grant No.~3491/21, 1982/22), the US-Israel Binational Science Foundation and US National Science Foundation, the Leona M.~and Harry B.~Helmsley Charitable Trust, the Shimon and Golde Picker - Weizmann Annual Grant, and the Laboratory in Memory of Leon and Blacky Broder.

\bibliographystyle{latexmkrc}
\bibliography{ProbabilisticCNOT}
\clearpage \onecolumngrid 
\renewcommand{\thesection}{S\arabic{section}}
\renewcommand{\theequation}{S\arabic{equation}}
\renewcommand{\thefigure}{S\arabic{figure}}
\renewcommand{\thetable}{S\arabic{table}}

\setcounter{section}{0}
\setcounter{equation}{0}
\setcounter{figure}{0}
\setcounter{table}{0}

\addcontentsline{toc}{section}{Supplementary Material}

\section*{Supplementary Material}
\section{Single photon source}

\noindent The photon source is based on a three-level ladder excitation scheme in thermal rubidium vapor, employing two pump fields at 780 nm and 776 nm to drive the transition  $\ket{5S_{1/2}, F=2}\rightarrow\ket{5P_{3/2}, F=3}\rightarrow\ket{5D_{5/2}, F=4}$. The excitation is two-photon resonant and detuned by -1.1 GHz from the intermediate $\ket{5P_{3/2}}$ level. The vapor cell has an optical depth (OD) of 4, yielding signal photons with a full width at half maximum (FWHM) of approximately 0.9-1.0 ns. 

Due to the symmetry of the vapor cell, spontaneous four-wave mixing produces signal and idler photons along phase-matched directions on either side of the optical axis in the horizontal plane. The resulting field modes are labeled i1 and i2 (idler), and s1 and s2 (signal), on either side of the optical axis, effectively operating as two independent photon-pair sources within the same vapor cell. Detection of an idler photon heralds the presence of its corresponding signal photon. 

\begin{figure}[!h]
  \centering
  \begin{minipage}[b]{0.48\textwidth}
    \includegraphics[width=0.9\textwidth]{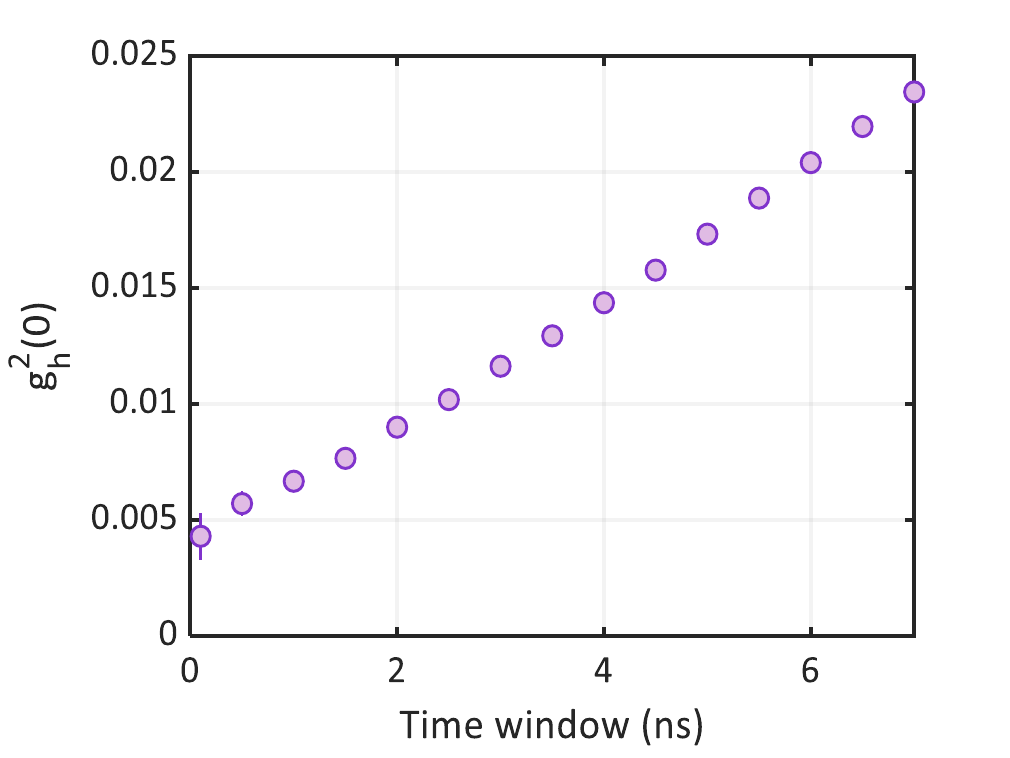}
    \caption{Heralded $g^{(2)}_\mathrm{h}(0)$ of the signal photon s1 generated by the source, as a function of coincidence window.}
    \label{fig:g2signal1}
  \end{minipage}
  \hfill
  \begin{minipage}[b]{0.48\textwidth}
    \includegraphics[width=0.9\textwidth]{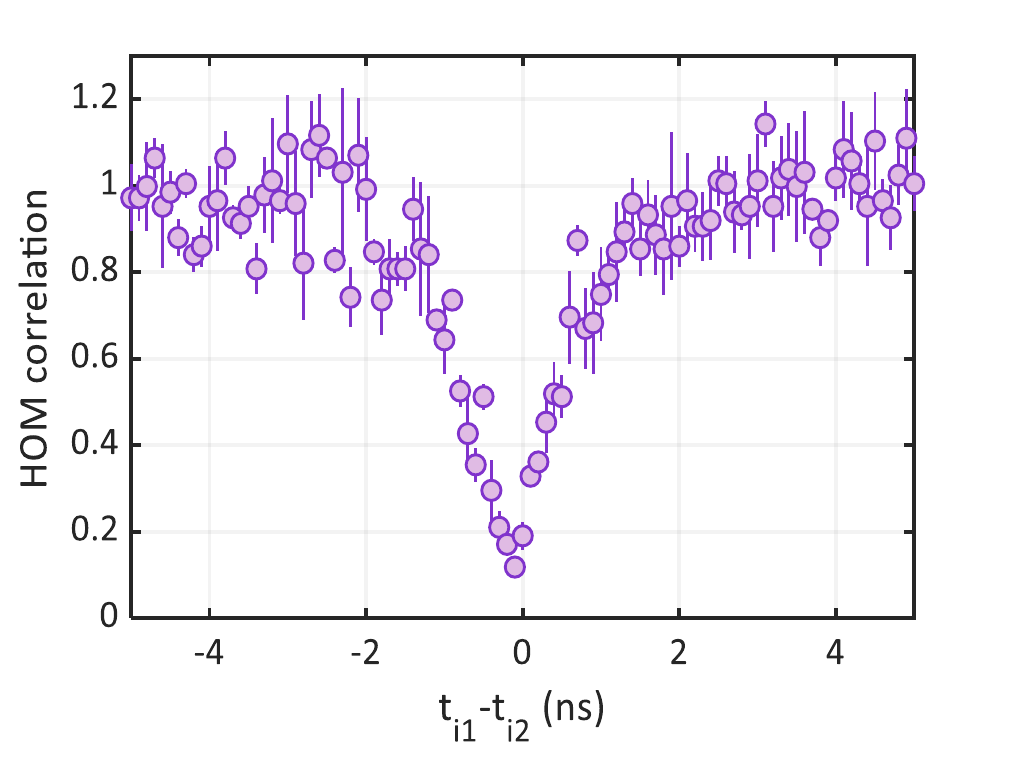}
    \caption{Hong-Hu-Mandel (HOM) interference measurement of the source's signal photons s1 and s2, plotted as a function of the time difference between the idler photons i1 and i2, with $\tau_\textrm{int}$ = 3.5 ns.}
    \label{fig:homsource}
  \end{minipage}
\end{figure}

The heralding efficiency is measured to be $18\%$ with $[g^{(2)}_\mathrm{s-i}]_{\mathrm{max}}=844$
in channel 1, and $22.7\%$ with $[g^{(2)}_\mathrm{s-i}]_{\mathrm{max}}=731$ in channel 2. Here $[g^{(2)}_\mathrm{s-i}]_{\mathrm{max}}$ is the maximum of the normalized signal-idler cross-correlation $g^{(2)}_\mathrm{s-i}$, often denoted as the signal-to-noise ratio of the source photons.
The single photon count rate can be tuned from $4\cdot 10^4$/sec to $4\cdot 10^5$/sec by adjusting the pump powers. In our experiment, we use 300 $\mu$W for the 780-nm pump and 1 mW for the 776-nm pump, resulting in a heralded single-photon generation rate of approximately $4\cdot 10^4$/sec. 
The single-photon nature of the source is confirmed via conditional autocorrelation measurements: the heralded autocorrelation of s1, conditioned on detection of i1, is found to be $g^{(2)}_\mathrm{h}(0)=0.0129(1)$ for coincidence window of 3.5 ns, see Fig.~\ref{fig:g2signal1}.
To assess the indistinguishability of the photons generated in the two source channels, we perform a Hong-Ou-Mandel (HOM) interference experiment. Signal photons s1 and s2 are interfered at a symmetric fiber beam-splitter via separate input ports, and the resulting coincidence rate is measured as a function of the temporal delay between the heralding idler detections, $t_{\mathrm{i}1} - t{\mathrm{i}2}$. Figure \ref{fig:homsource} shows the resulting HOM dip for photon pairs generated directly from the source. For the analysis, we define an integration window of duration $\tau_\textrm{int}$ that specifies which four-photon detection events are considered valid double-heralded single-photon pairs. Specifically, we include events where either $t_\mathrm{s_a}-t_\textrm{i1}$, $t_\mathrm{s_b}-t_\textrm{i2}$ or $t_\mathrm{s_a}-t_\textrm{i2}$, $t_\mathrm{s_b}-t_\textrm{i1}$ are both within the chosen $\tau_\textrm{int}$. Here, $t_\mathrm{s_a}$ and $t_\mathrm{s_b}$ denote the detection times of the signal photons at the two output ports of the beam splitter. 
Further information on the photon source architecture and performance can be found in Refs.~\cite{Davidson_2021,davidson2023newsource}.

\section{Quantum memory}
\noindent The quantum memory operates on a three-level ladder scheme based on the FLAME protocol \cite{Finkelstein2018}, ensuring compatibility between the photons generated by the source and those stored in the memory. A strong 776-nm drive field, detuned by $-60$ MHz from the intermediate state $\ket{5P_{3/2}}$, is used for both storage and retrieval. The drive pulses are carved from a continuous-wave (CW) field using a pair of Pockels cells, producing 2-ns-long pulses at a maximum duty cycle of 1:5000. The pulses reach a peak power of approximately 1 W. A digital delay generator triggers the Pockels cells upon detection of the heralding idler photons, such that the first pulse stores the signal photon s1 in the memory, and the second pulse retrieves it in synchrony with s2. 

Optical pumping is implemented using annular pump and repump beams to prepare the atoms in the stretched spin state $\ket{5S_{1/2}, F=2, m_{F}=2}$. These beams propagate at a small angle relative to the drive field. Additional details on the FLAME protocol and experimental implementation are provided in Ref.~\cite{Davidson2023}. 

\begin{figure}[!h]
  \centering
  \begin{minipage}[b]{0.48\textwidth}
    \includegraphics[width=0.85\textwidth]{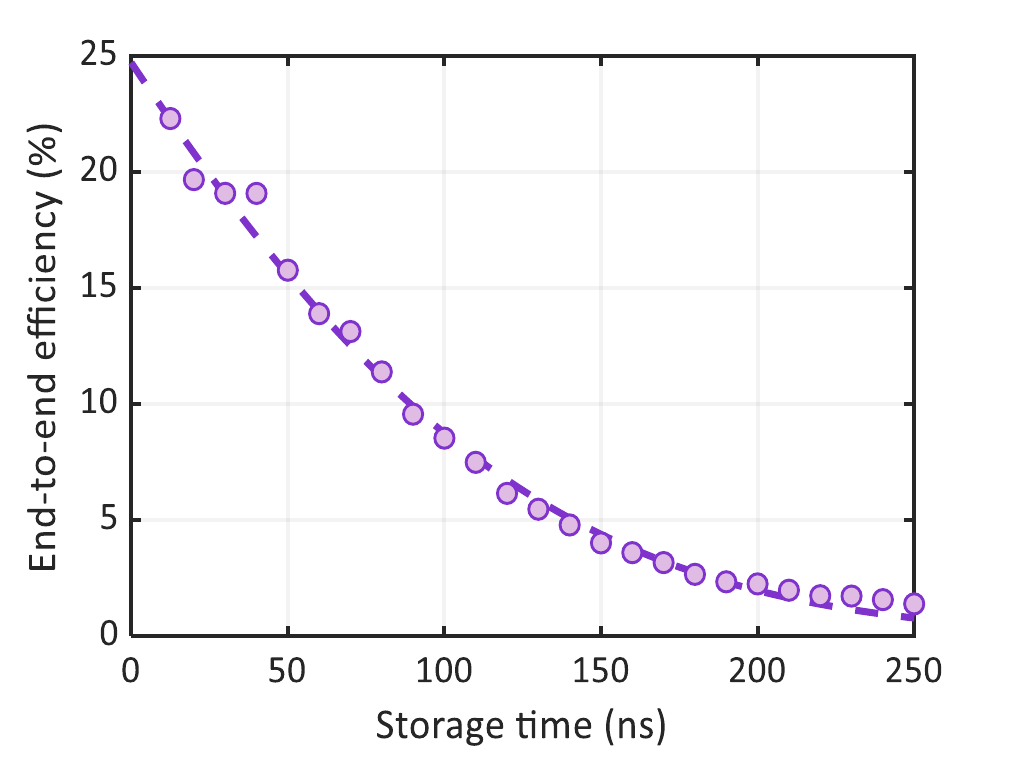}
    \caption{End-to-end quantum memory efficiency for storage and retrieval of a single photon, as a function of storage time.}
   \label{fig:memoryeff}
  \end{minipage}
  \hfill
  \begin{minipage}[b]{0.48\textwidth}
    \includegraphics[width=0.85\textwidth]{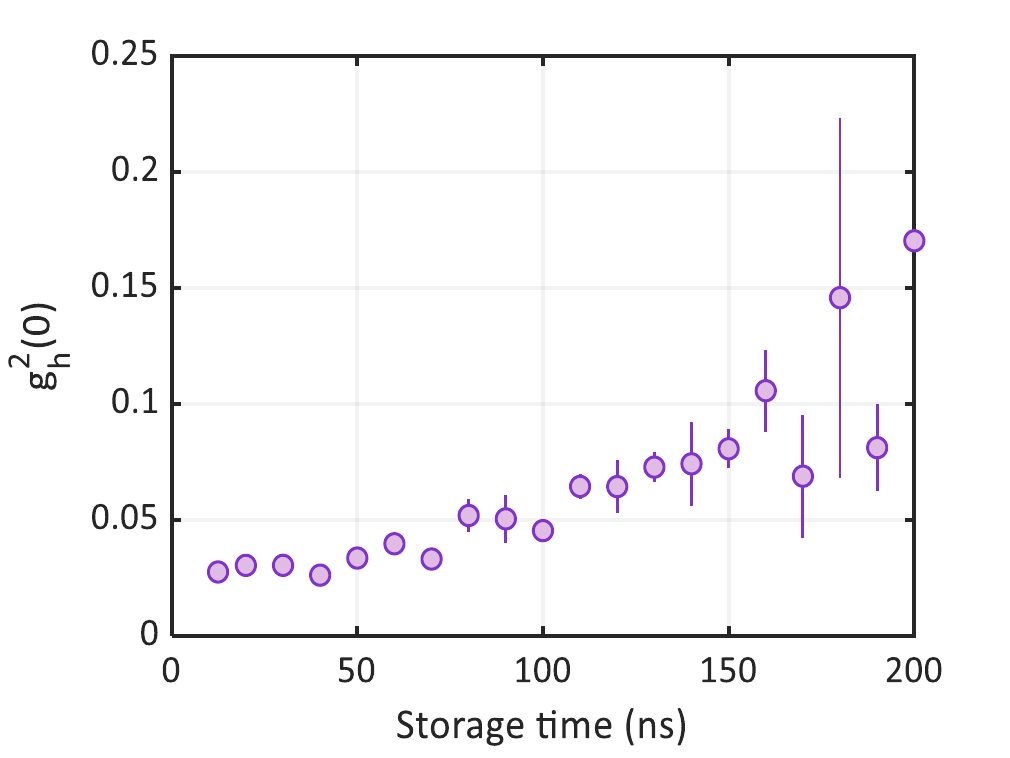}
    \caption{Autocorrelation $g^{(2)}_\mathrm{h}(0)$ of the retrieved photon from the memory as a function of storage time, for $\tau_\textrm{int}= 3.5$~ns.}
    \label{fig:g2memory}
  \end{minipage}
\end{figure}

The retrieved photon has a temporal width (FWHM) of approximately 1.5 ns. The end-to-end efficiency of the memory, including optical coupling losses, is $25 \%$ at zero storage time (Fig.~\ref{fig:memoryeff}).
To verify the single-photon nature of the retrieved photon, we measure its autocorrelation $g^{(2)}_\mathrm{h}(0)$ as a function of the storage time (Fig.~\ref{fig:g2memory}). For a storage time of 10 ns, we obtain $g^{(2)}_\mathrm{h}(0)=0.028(5)$, which is only slightly above the source value and still satisfies $g^{(2)}_\mathrm{h}(0) \ll 1$. This confirms that the quantum character of the photon is preserved through the storage and retrieval process.

\section{Photon synchronization}
\noindent Photon synchronization is achieved by integrating the photon source with the quantum memory and employing a digital electronic scheme to trigger the storage and retrieval pulses. A pair of digital delay generators ensure that the Pockels cells generate the drive pulses only when both idler photons are detected within a 100-ns coincidence window. Since the Pockels cells have a post-operation dead time of 1.5~$\mu$s, this electronic gating ensures that they are triggered only by dual-idler events, while additional triggers are ignored during the pulse generation period. Further details on the synchronization protocol and experimental implementation can be found in Ref.~\cite{Davidson_2023_prl} and its Supplementary Material.

Photon synchronization increases the rate of usable pairs by a factor of $19.1$ compared to the rate of accidental pairs from the source when operating at a single-photon generation rate of 40 kHz. This enhancement factor remains nearly constant across different choices of the integration window duration $\tau_\textrm{int}$.  The integration window excludes the initial rise of the photon pulses.

\begin{figure}[t]
\includegraphics[width=0.44\textwidth]{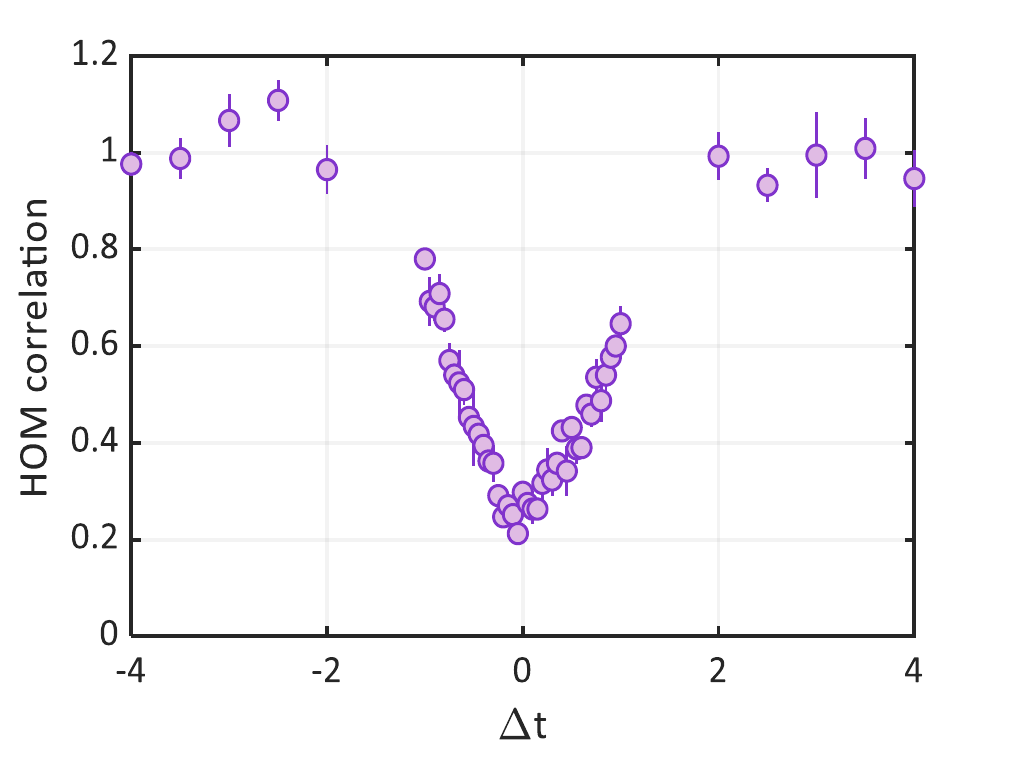}
\parbox{0.6\linewidth}{\caption{HOM interference of synchronized photons as a function of the programmed delay between s2 and s1'. Here, $\tau_\textrm{int}$ = 3.5 ns.}\label{fig:homsync}} 
\end{figure}

We perform the HOM interference measurement between the source and memory photons to quantify their indistinguishability. In this measurement, the retrieved photon s1$'$ and the source photon s2 are interfered at a symmetric fiber beam splitter, and the coincidence rate is measured as a function of $\Delta t$. The parameter $\Delta t$ defines the programmed delay between the expected arrival time of s2 (from the source) and the retrieval time of s1' (from the memory). For synchronized photon pairs, the double-heralding condition for data analysis requires that either $t_\mathrm{s_a}-t_\textrm{i2}$, $t_\mathrm{s_b}-(t_\textrm{i2}+\Delta t)$ or $t_\mathrm{s_a}-(t_\textrm{i2}+\Delta t)$, $t_\mathrm{s_b}-t_\textrm{i2}$ both fall within $\tau_\textrm{int}$.
As shown in Fig.~\ref{fig:homsync}, the HOM visibility for the synchronized photon pair is reduced compared to that of accidental photon pairs. This reduction arises from the mismatch in temporal profiles between the source and memory photons (Fig.~1d in the main text). In particular, the retrieved photon is substantially reshaped by the memory, evolving from an intially asymmetric to a more symmetric temporal waveform.

\section{CNOT gate operation}

\begin{figure}[b]
    \centering
    \includegraphics[width=0.7\columnwidth]{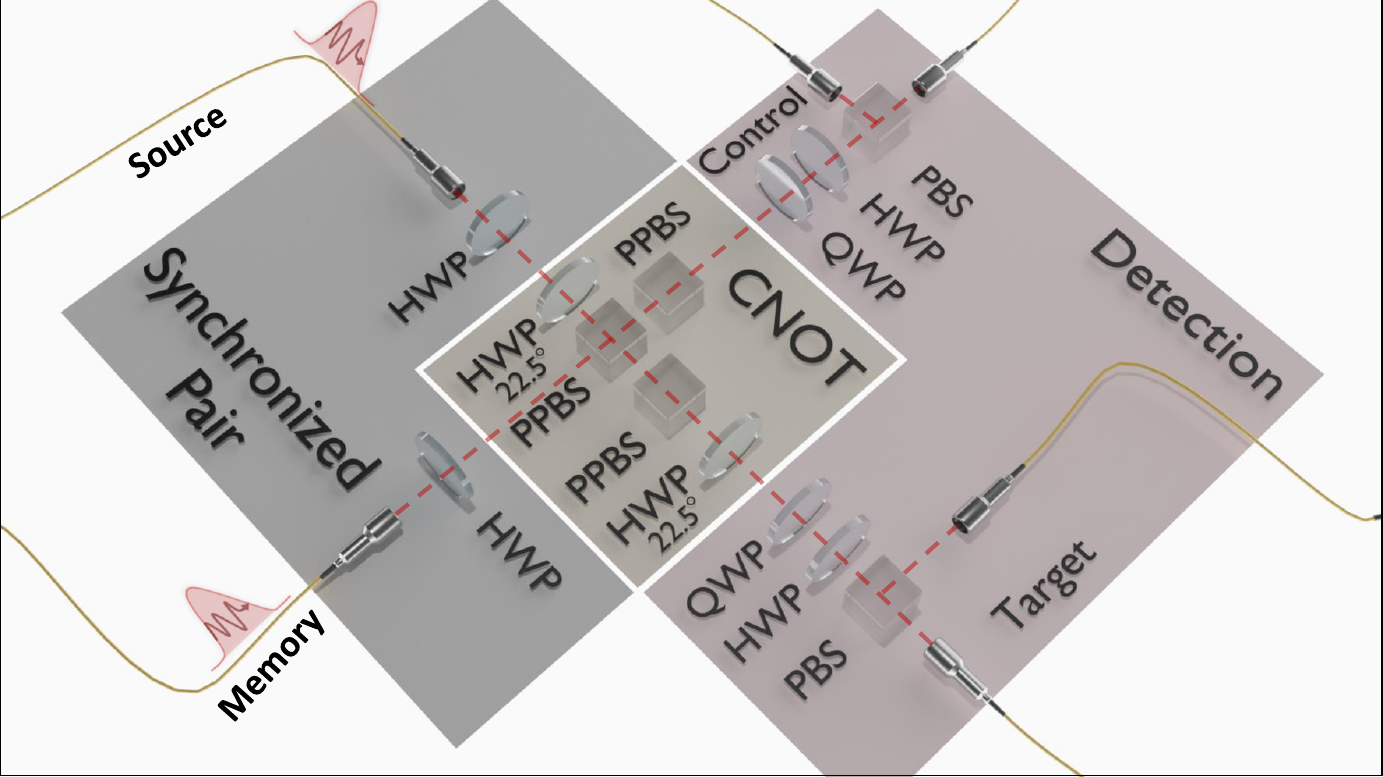}
    \caption{ \textbf{Experimental scheme for a KLM-type CNOT quantum gate}. The input qubit states are prepared using half-wave plates (HWPs). The gate consists of three partially polarizing beam splitters (PPBSs) and two HWPs, which are set to $22.5^\circ$ to implement Hadamard operations. At the output, the qubits pass through HWPs and quarter-wave plates (QWPs), followed by polarizing beam splitters (PBS) and single-photon detectors, enabling full tomographic state analysis.}
    \label{fig:CNOt_System}
\end{figure}

\begin{figure}[tb]
    \centering
    \includegraphics[width=0.5\columnwidth]{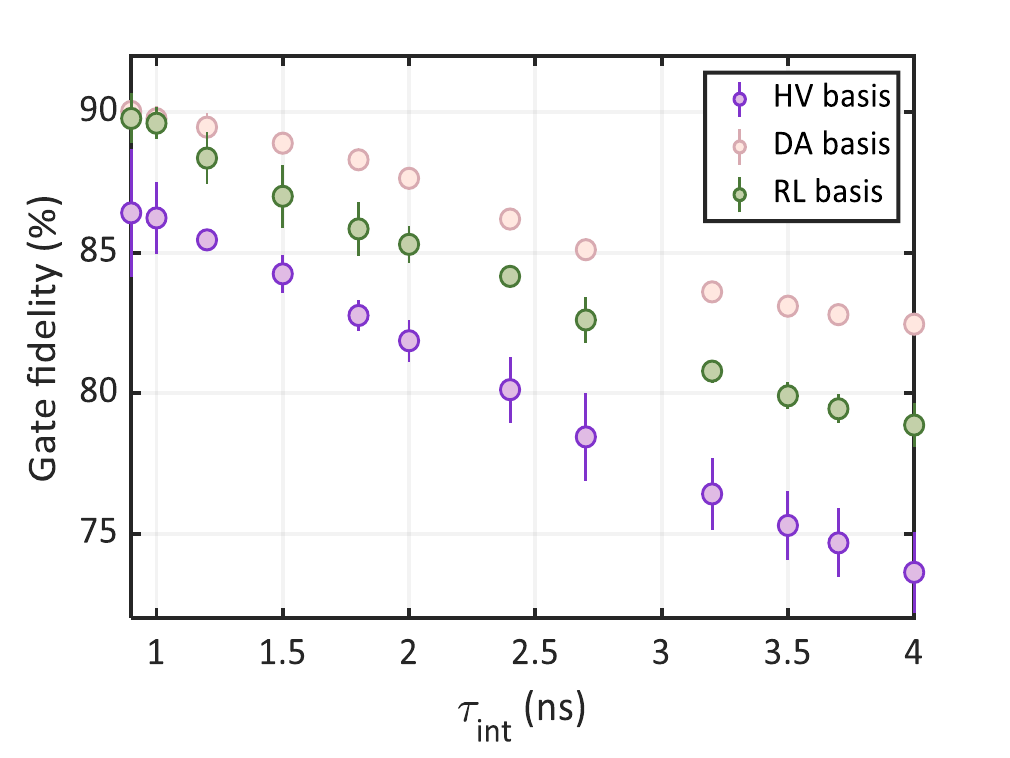}
    \caption{Truth table fidelity as a function of $\tau_\textrm{int}$ for the three polarization basis.}
    \label{fig:truthtablefidgate}
    \vspace{1cm}
\end{figure}

\noindent The CNOT gate setup is shown and described in Fig.~\ref{fig:CNOt_System}. When the two photons are synchronized, the retrieval delay $\Delta t$ is set to zero. The only parameter varied in our data analysis is the integration window $\tau_\textrm{int}$, which controls the temporal overlap---and hence the indistinguishability---of the photons. Coincidence counts between detector pairs are normalized to the total number of coincidences across all four detectors, effectively defining the photonic waveform within the integration window. 
We evaluate the gate performance across three polarization bases as a function of $\tau_\textrm{int}$, as shown in Fig.~\ref{fig:truthtablefidgate}. As the integration window is narrowed, the temporal indistinguishability of the photons and the SNR improve, leading to an increase in gate fidelity.

We perform full quantum state tomography by conducting 144 measurements on the output state, using four detectors (two per output qubit) and 36 distinct measurement settings (six per output state). When the input qubits are prepared in the states $\ket{\mathrm{AV}}$, $\ket{\mathrm{DV}}$, $\ket{\mathrm{AH}}$, and $\ket{\mathrm{DH}}$, the action the CNOT circuit ideally generates the four Bell states. 
The corresponding output density matrices, presented in Fig.~2b in the main text, are reconstructed using a maximum likelihood algorithm following Refs.~\cite{altepeter2005photonic,james2001measurement}.

\begin{figure}[tb]
    \centering
    \includegraphics[width=\columnwidth]{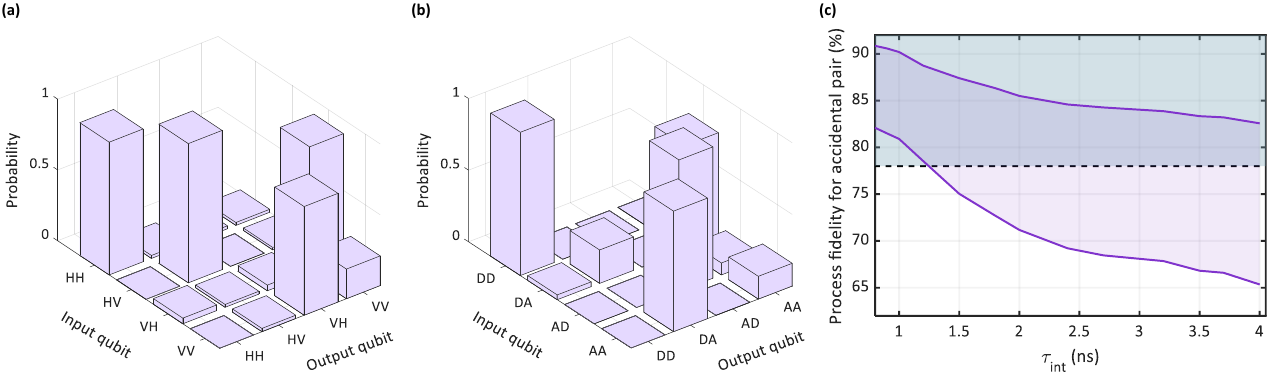}
    \caption{\textbf{Gate performance with accidental (unsynchronized) photon pairs. (a)} Truth table in the HV basis. \textbf{(b)} Truth table in the DA basis. \textbf{(c)} Upper and lower bounds on the process fidelity (solid purple lines). The dashed line draws the threshold above which Bell's inequality breaks.}
    \label{fig:truthtablesource}
\end{figure}

We also characterize the performance of the CNOT gate using accidental photon pairs generated directly from the source. 
For this configuration, we only take truth table measurements and the fidelity is found out to be $F_\textrm{ZZ}=0.902(8)$ and $F_\textrm{XX}=0.907(8)$ for $\tau_\textrm{int}=1$ ns (Fig.~\ref{fig:truthtablesource}a,b). From the truth table measurements, we calculate the upper and lower bounds on the process fidelity  (Fig.~\ref{fig:truthtablesource}c), since full quantum state tomography with accidental pairs would require prohibitively long measurement times (exceeding 150 hours).

\newpage

\section{\label{theory_vis_fid} Theoretical calculation for the dependence of interference-based CNOT gates on the photon indistinguishability. }

\begin{figure}[b]
    \centering
    \includegraphics[width=0.75\columnwidth]{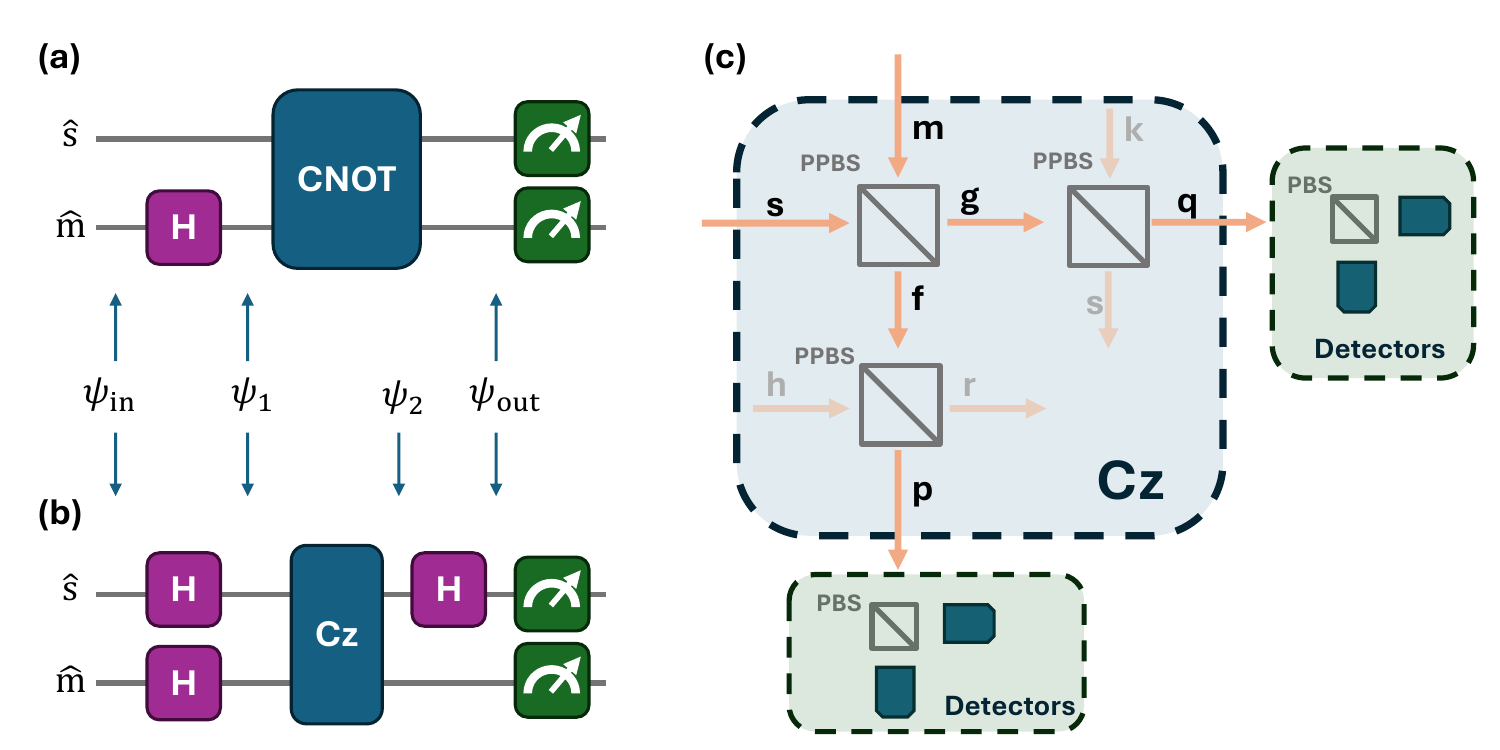}
    \caption{\textbf{(a)} Circuit representation of the CNOT gate, with the control and target qubits denoted as `m' (for `memory') and `s' (for `source'), respectively. An additional Hadamard transformation is applied to the control qubit to generate Bell states. \textbf{(b)} An equivalent circuit, utilized in the theoretical derivation, consisting of a controlled-Z gate sandwiched between two Hadamard gates on the target qubit. \textbf{(c)} Configuration of the PPBS in mode notations.}
    \label{fig:PPBS}
\end{figure}

\noindent We derive a general relation between the performance of an interference-based CNOT gate and the temporal indistinguishability of the input photons. The gate architecture and mode labeling are shown in Fig.~\ref{fig:PPBS}. 
We assume an idealized setup with no imperfections in the transmission and reflection coefficients of the PPBSs: $T_\mathrm{H} = 1$, $R_\mathrm{H} = 0$, $T_\mathrm{V} = \frac{1}{3}$, and $R_\mathrm{V} = \frac{2}{3}$. 

The mode transformation by the first PPBS, where the HOM (two-photon) interference occurs, is given by
\begin{align} \label{1st_PPBS}
    \hat{m}_\mathrm{H}^{\dagger} &= \hat{f}_\mathrm{H}^{\dagger} \quad \quad \quad 
    \hat{m}_\mathrm{V}^{\dagger} 
=\left(\sqrt{T_\mathrm{V}}\hat{f}_\mathrm{V}^{\dagger}+i\sqrt{R_\mathrm{V}}\hat{g}_\mathrm{V}^{\dagger}\right) =\frac{1}{\sqrt{3}}\left(\hat{f}_\mathrm{V}^{\dagger}+i\sqrt{2}\hat{g}_\mathrm{V}^{\dagger}\right) \notag, \\
    \hat{s}_\mathrm{H}^{\dagger}&= \hat{g}_\mathrm{H}^{\dagger}, \quad\quad\quad
    \hat{s}_\mathrm{V}^{\dagger}= \left(\sqrt{T_\mathrm{V}}\hat{g}_\mathrm{V}^{\dagger}+i\sqrt{R_\mathrm{V}}\hat{f}_\mathrm{V}^{\dagger}\right) = \frac{1}{\sqrt{3}}\left(\hat{g}_\mathrm{V}^{\dagger}+i\sqrt{2}\hat{f}_\mathrm{V}^{\dagger}\right)  ,
\end{align}
where $\hat{x}_\mathrm{P}^{\dagger}$ creates a photon in mode $x\in$\{s,m,f,g,p,q\} with polarization P$\in$\{H,V\}. 
The second and third PPBSs have the same transmission and reflection coefficients, $T'_\mathrm{H}=T''_\mathrm{H}=1$ and $T'_\mathrm{V}=T''_\mathrm{V}=\frac{1}{3}$, and serve to balance the output amplitudes for all combinations of input photon polarizations. Due to the post-selection on events in which one photon is detected in each of the monitored output ports `p' and `q' of the PPBSs, we consider only these ports in the subsequent PPBSs' transformations, 
\begin{align}\label{2nd_PPBS}
    \hat{f}_\mathrm{H}^{\dagger} &= \sqrt{T'_\mathrm{V}}\hat{p}_\mathrm{H}^{\dagger}=\frac{1}{\sqrt{3}}\hat{p}_\mathrm{H}^{\dagger} \notag , \quad\quad\quad
    \hat{f}_\mathrm{V}^{\dagger}=\sqrt{T'_\mathrm{H}}\hat{p}_\mathrm{V}^{\dagger}=\hat{p}_\mathrm{V}^{\dagger} \notag, \\
    \hat{g}_\mathrm{H}^{\dagger}&= \sqrt{T''_\mathrm{V}}\hat{q}_\mathrm{H}^{\dagger}=\frac{1}{\sqrt{3}}\hat{q}_\mathrm{H}^{\dagger},\quad\quad\quad
    \hat{g}_\mathrm{V}^{\dagger}=\sqrt{T''_\mathrm{H}}\hat{q}_\mathrm{V}^{\dagger}=\hat{q}_\mathrm{V}^{\dagger}.
\end{align}

As a representative case, we consider the generation of the Bell state $\ket{\Phi^+}$ from the input state $\ket{\mathrm{DH}}_\mathrm{m,s}$, where the modes `m' and `s' correspond to the control and target qubits (memory and source photons), respectively. We represent this input state by
\begin{equation}
\left|\psi_\mathrm{in}\right\rangle = \frac{1}{\sqrt{2}} \iint  dt_\mathrm{m} dt_\mathrm{s} u_\mathrm{m}(t_\mathrm{m}) u_\mathrm{s}(t_\mathrm{s}) [\hat{m}_\mathrm{H}^{\dagger}(t_\mathrm{m})+\hat{m}_\mathrm{V}^{\dagger}(t_\mathrm{m})] \hat{s}_\mathrm{H}^{\dagger}(t_\mathrm{s})\left|00\right\rangle _\mathrm{m,s}\equiv \frac{1}{\sqrt{2}}f(t_\mathrm{m},t_\mathrm{s})\left[\ket{\mathrm{HH}}_\mathrm{m,s}+\ket{\mathrm{VH}}_\mathrm{m,s}\right]
    \label{psi_in_SI},
\end{equation}
where we defined the temporal profile of the two-photon wavefunction $f(t,t')=u_\mathrm{m}(t)u_\mathrm{s}(t')$  and adopted the shorthand notation $f(t_x,t_y) \ket{PQ}_{x,y} \equiv \int dt~ u_x(t) x^\dagger_P\int dt'~ u_y(t') y^\dagger_Q|0\rangle$. The time integrals here and throughout are taken over the chosen integration window, and the temporal profiles $u_x(t)$ are normalized such that $\int dt |u_x(t)|^2=1$.

After the first Hadamard transformation inside the gate, the state vector becomes
\begin{equation}
    \ket{\psi_\mathrm{in}} \xrightarrow{\text{Hadamard}}\ket{\psi_1}=\frac{1}{2}f(t_\mathrm{m},t_\mathrm{s})
    (\ket{\mathrm{HH}}_\mathrm{m,s}+\ket{\mathrm{HV}} _\mathrm{m,s}+\ket{\mathrm{VH}}_\mathrm{m,s}+\ket{\mathrm{VV}}_\mathrm{m,s})
    \label{psi_1}.
\end{equation}
Then, following the transformations in Eqs.~(\ref{1st_PPBS}) and (\ref{2nd_PPBS}) that realize a CZ gate, we get
\begin{align}
        \ket{\psi_1} \xrightarrow{\text{PPBSs}}\ket{\psi_2}= \sqrt{2}C\iint dt_\mathrm{p} dt_\mathrm{q} f(t_\mathrm{p},t_\mathrm{q})  
        (\ket{\mathrm{HH}}_\mathrm{p,q}+\ket{\mathrm{HV}} _\mathrm{p,q}+\ket{\mathrm{VH}}_\mathrm{p,q}+\ket{\mathrm{VV}}_\mathrm{p,q})- 2f(t_\mathrm{q},t_\mathrm{p})\ket{\mathrm{VV}}_\mathrm{p,q}
        \label{psi_2}.
\end{align}

where $C$ is a normalization constant needed after the post-selection. 
The last Hadamard transformation produces the output state
\begin{align}
        \ket{\psi_2} \xrightarrow{\text{Hadamard}}\ket{\psi_\mathrm{out}}= C
        [f(t_\mathrm{p},t_\mathrm{q})\left|\mathrm{HH}\right\rangle_\mathrm{p,q} +2f_{-}(t_\mathrm{p},t_\mathrm{q})\left|\mathrm{VH}\right\rangle_\mathrm{p,q}+f(t_\mathrm{q},t_\mathrm{p})\left|\mathrm{VV}\right\rangle_\mathrm{p,q}]
        \label{psi_out_SI}.
\end{align}
where $f_{\pm}(t,t')=[u_\mathrm{m}(t)u_\mathrm{s}(t') \pm u_\mathrm{m}(t')u_\mathrm{s}(t)]/2=[f(t,t')\pm f(t',t)]/2$. 

We notice that
\begin{align}
\iint dt dt'f^{*}(t,t')f(t,t')=1  \quad \mathrm{and} \quad
\iint dt dt'f^{*}_{\pm}(t,t')f_{\pm}(t,t')=\frac{1\pm \eta}{2},
\end{align}
where 
\begin{equation}
\eta=\left|\int dt u^*_\mathrm{m}(t) u_\mathrm{s}(t) \right|^2
\end{equation}
is the overlap integral of the two photon waveforms. Normalization of $\ket{\psi_\mathrm{out}}$ therefore requires $C = 1/\sqrt{4-2\eta}$.

In the main text, we derived the state generation fidelity using a decomposition into Bell-like states with explicit two-photon temporal wavefunctions. Here we present an alternative approach based on tracing over the temporal degrees of freedom to construct the reduced density matrix $\rho=\iint dt dt' |\psi_\mathrm{out}\rangle\langle\psi_\mathrm{out}|$,
\begin{align}\label{rho_phip}
\rho_{\Phi^{+}} (\eta) = \frac{1}{4-2\eta}\cdot 
        \begin{bmatrix}
              1           & 0 & 1-\eta &      \eta          \\
                   0                 & 0 &             0            &           0                  \\
        1-\eta     & 0 &           1-\eta         &  1-\eta    \\
              \eta         & 0 & 1-\eta &        1           \\
        \end{bmatrix}.
\end{align}
Notably, $\rho$ does not take the form of a Werner state $F'\ket{\Phi^+}\bra{\Phi^+} + (1-F')(\mathbb{I}/4)$, and hence captures more structure than common phenomenological models \cite{RiedelGarding2021Jun}. The fidelity of the generated state is calculated using  $F = \text{Tr}(\rho \rho_{\text{ideal}}) + 2\sqrt{\det(\rho)\det(\rho_{\text{ideal}})}$, where $\rho_{\text{ideal}}=\ket{\Phi^+}\bra{\Phi^+}$ \cite{jozsa1994fidelity}. Since $\rho_{\text{ideal}}$ is pure, this expression simplifies to $F = \text{Tr}(\rho \rho_{\text{ideal}})$. We then obtain
\begin{align}
F(\eta) =  \frac{1}{2}\cdot \frac{1+\eta}{2-\eta},
     \label{Fidelity}
\end{align}
in agreement with the result derived in the main text [Eq.~(5)].  

A similar procedure can be applied to input states $\ket{\mathrm{HV}}_{\mathrm{m,s}}$, $\ket{\mathrm{VH}}_{\mathrm{m,s}}$, and $\ket{\mathrm{VV}}_{\mathrm{m,s}}$ to obtain the reduced density matrices corresponding to the other Bell states, 
\begin{align}
\rho_{\Phi^{-}} (\eta) = \frac{1}{4-2\eta}\cdot 
        \begin{bmatrix}
              1            & 0 & 1-\eta &      -\eta          \\
                   0                 & 0 &             0            &           0                  \\
        1-\eta     & 0 &           1-\eta         &  1-\eta    \\
              -\eta         & 0 & 1-\eta &        1           \\
        \end{bmatrix}
\quad\quad \mathrm{and} \quad\quad
\rho_{\Psi^{\pm}} (\eta) = \frac{1}{4-2\eta}\cdot 
        \begin{bmatrix}
        0 &          0                  &          0                  &                       0               \\
        0 &     1             &    \pm\eta          &           1-\eta           \\
        0 &   \pm\eta           &    1              &          1-\eta    \\
        0 &    1-\eta &    1-\eta &                   1-\eta           \\
        \end{bmatrix}.
     \label{rho_ideal_psi_p}
\end{align}
which yields the same expression for the fidelity as in the main text [Eq.~(5)].

\end{document}